\documentclass[10pt,letterpaper]{article}
\usepackage{opex3}

\begin{document}


\title{Nonlinear optical contrast enhancement for optical coherence tomography}

\author{Claudio Vinegoni, Jeremy S. Bredfeldt, and Daniel L. Marks}

\address{Beckman Institute for Advanced Science and Technology, University of Illinois at Urbana-Champaign}

\author{Stephen A. Boppart}

\address{Department of Electrical and Computer Engineering\\
Beckman Institute for Advanced Science and Technology,
Department of Bioengineering \\ College of Medicine\\
University of Illinois at Urbana-Champaign, 405 N. Mathews Ave.,
Urbana, IL 61801}

\email{boppart@uiuc.edu} 

\homepage{http://nb.beckman.uiuc.edu/biophotonics/} 


\begin{abstract}
We present a new interferometric technique for measuring Coherent
Anti-Stokes Raman Scattering (CARS) and Second Harmonic Generation
(SHG) signals. Heterodyne detection is employed to increase the
sensitivity in both CARS and SHG signal detection, which can also
be extended to different coherent processes. The exploitation of
the mentioned optical nonlinearities for molecular contrast
enhancement in Optical Coherence Tomography (OCT) is presented.
Numerical simulations for both coherent nonlinear processes are
performed in order to determine the properties of the signal
expected at the exit of the described nonlinear interferometer.
\end{abstract}





\section{Introduction}

Optical coherence tomography (OCT) is an emerging biomedical
imaging technology that has been applied to a wide range of
biological, medical, and materials investigations. OCT was first
developed in the early 1990s for noninvasive imaging of biological
tissue \cite{huang1991} and is capable of imaging tissue
microstructures at near histological resolutions
\cite{boppart1998}. Axial resolution of 10 $\mu$m is common for
standard OCT, where for ultrahigh resolution OCT, an axial
resolution in the 1 $\mu$m range has been recently achieved using
broadband-continuum generation from a photonic crystal fiber
\cite{wang2003}.

The advantages of OCT  in biomedical imaging, compared to other
imaging techniques,  are quite numerous \cite{fujimoto2003}. In
particular, OCT can provide imaging resolutions that approach
those of conventional histopathology and can be performed {\it in
situ} \cite{fujimoto2003}. Despite its advantages, a serious
drawback to OCT is that the linear scattering properties of
pathological tissue probed by OCT are often morphologically and/or
optically similar to the scattering properties of normal tissue.
For example, although morphological differences between normal and
neoplastic tissues may be obvious at later stages of tumor
development, it is frequently difficult to optically detect
early-stage tumors \cite{lee2003}.

This implies a need for novel contrast enhancing mechanisms for
OCT. Examples of methods that have been recently developed,
include: spectroscopic OCT \cite{morgner2000}, pump and probe
techniques \cite{rao2003}, and the use of engineered microspheres
\cite{lee2003} or microbubbles \cite{barton2002}. Spectroscopic
OCT (SOCT) measures the spectral absorption from tissues by
measuring the spectral differences between the source and the
backscattered interference signal to provide information about the
properties of the scatterers in the sample. However, this
technique is limited to the identification of scatterers that have
absorption within the bandwidth of the optical source. Pump and
probe contrast enhancement for OCT imaging relies on transient
absorptions in the sample under investigation that are induced by
an external pump beam. Unfortunately it is necessary in most cases
to introduce different contrast agents depending on the excitation
source and on the transient spectra of the molecules under
investigation \cite{rao2003}. An alternative way to obtain
contrast enhancement in OCT includes the use of exogenous contrast
agents such as engineered microspheres. These microspheres can be
targeted to cell receptors and change the optical scattering or
absorption characteristics in selected regions, providing
molecular specific contrast \cite{lee2003}.

As clearly evidenced above, most of the current methods  (if not
all)  that are currently used  to enhance contrast imaging in OCT
require the presence of a contrast agent, which can modify the
biology under investigation. It follows there is a need for new
techniques that could help eliminate this limitation. In this
paper, we propose and present new methods to achieve enhanced OCT
contrast, exploiting optical nonlinearities. The nonlinear effects
on which we focus in this work in particular are  Coherent
Anti-Stokes Raman Scattering (CARS) and Second Harmonic Generation
(SHG), but the general idea could be easily extended to other
nonlinear effects such as Third Harmonic Generation (THG),
Coherent Stokes Raman Spectroscopy (CSRS), and stimulated emission
in active materials (i.e. InGaAs, GaAs based materials, etc.).

\section{CARS contrast enhancement}
\subsection{Theory}
\label{carstheory}
 It is  well known that the nonlinear
polarization for a material can be expressed as a function of the
incident electric field vector $\bar{E}$:
%
%
\begin{equation}
\label{expansionp}
 \bar{P} = \epsilon_0 \left( \chi^{(1)}\cdot \bar{E} + \chi^{(2)}
: \bar{E}\bar{E} + \chi^{(3)} \, \vdots \, \bar{E}\bar{E}\bar{E} +
\ldots \right)
\end{equation}
with $\bar{P}$ the induced polarization, $\chi^{(n)}$ the n-th
order nonlinear susceptibility, and $\epsilon_0$ the vacuum
permittivity.  This implies that for high intensities (i.e. in a
nonlinear regime) the induced polarization is no longer directly
proportional to the incoming electric field vector $\bar{E}$.
Usually the first term $\chi^{(1)}$ represents the main
contribution to $\bar{P}$ and  describes linear effects as
absorption or reflection. The second term  is responsible for
nonlinear effects like SHG and sum-frequency generation which  we
will consider in Sect. \ref{sectionshg}. The third term
$\chi^{(3)}$ is responsible for phenomena involving four photons,
like CARS, four wave mixing (FWM), third harmonic generation
(THG), and nonlinear refraction.

In this section we focus on CARS, a spectroscopic technique that
has recently received increasing attention for its applications
for vibrational imaging \cite{zumbusch1999}. In CARS spectroscopy,
the frequencies of two incident lasers, $\omega_p$ and  $\omega_s$
(Pump and Stokes, respectively), are selected such that the
difference in frequency $\omega_p$-$\omega_s$= $\omega_v$ is equal
to the frequency of a  Raman-active vibrational mode present in
the molecule under study \cite{dermtroeder1998}.  As evidenced
from Eq.\ref{expansionp}, CARS is a nonlinear, four-wave mixing
process. It follows the CARS field is a result of the interaction
between four photons and is generated in the phase-matching
direction at the anti-Stokes frequency $\omega_{AS}=2\omega_p
-\omega_s$, implying that the CARS signal intensity is linearly
dependent on the Stokes field intensity and
 quadratically dependent on the pump field intensity.
%
%
\begin{figure}[htbp]
\centering\includegraphics[width=7cm]{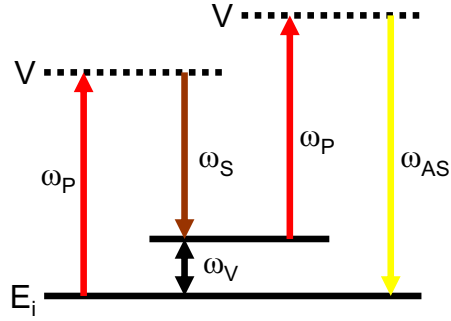}
\caption{Simplified energy diagram for CARS. $\omega_P$ indicates
the pump laser frequency, $\omega_s$ the Stokes laser frequency,
$\omega_{AS}$ the Anti-Stokes frequency (emitted CARS signal), and
$\omega_v$ the Raman frequency corresponding to an active
vibrational transition.} \label{transitioncars}
\end{figure}
Note that CARS is a coherent process, with the phase of the
anti-Stokes field related to the phase of the excitation field.
Therefore, constructive interference of the anti-Stokes field
causes the CARS signal to be significantly larger than the
spontaneous Raman signal, given the same average excitation power
\cite{cheng2002bis}. All these characteristics have allowed CARS
to be successfully employed to provide vibrational contrast in
scanning microscopy
\cite{zumbusch1999,cheng2002bis,duncan1982,wurpel2002}. CARS
microscopy generally involves scanning overlapped and tightly
focused pump and Stokes lasers through a sample while measuring
the anti-Stokes signal amplitude point by point
\cite{hashimoto2000}. The first CARS microscope \cite{duncan1982}
used non-collinear pump and Stokes visible lasers to prove
microscopic imaging of the spatial distribution of deuterated
molecular bonds in a sample of onion skin.

Picosecond lasers were used by Hashimoto et
al.\cite{hashimoto2000} to improve the Raman spectral resolution
and to further reduce the non-resonant background signal. Cheng et
al.\cite{cheng2002bis} theoretically evaluated the use of CARS in
microscopy and recently used polarization, epi-directional,
counter propagating, and forward CARS microscopy to study cell
biology. Multiplex CARS imaging has also been demonstrated to
provide contrast based on one or more vibrational spectral
features \cite{wurpel2002}.

In each of these CARS microscopy techniques, the anti-Stokes
photons are counted in order to estimate the density of the Raman
scatterers in the focal volume of the microscope. Unfortunately,
the spectral phase information is lost in this process. In this
paper we present a new interferometric technique called Nonlinear
Interferometric Vibrational Imaging (NIVI) \cite{marks2003} with
the capability for heterodyne detection and the possibility to
obtain a full reconstruction of the magnitude and phase of the
sample Raman susceptibility \cite{marks2003bis,bredfeldt2003}.

\subsection{Experimental Setup}

\begin{figure}[htbp]
\centering\includegraphics[width=9cm]{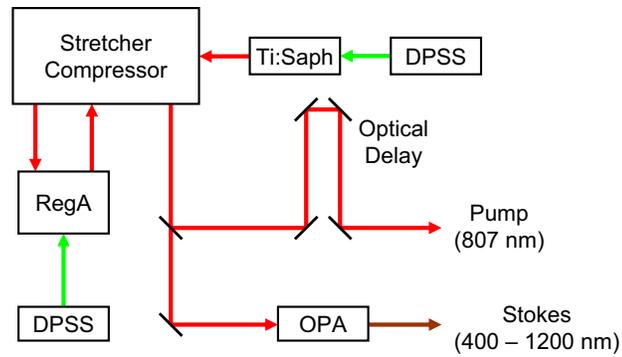} \caption{Setup
used to generate the Stokes and the Pump excitation fields. }
\label{setuplasers}
\end{figure}

The laser system constructed to create the laser fields necessary
to stimulate the CARS signal in  the samples is shown in
Fig.\ref{setuplasers}.
%
%
A diode pumped frequency doubled Nd:YVO$_4$ laser (Coherent,
Verdi) was used to pump a mode-locked Ti:sapphire oscillator
(KMLabs) operating at a center wavelength of 807 nm, with a
bandwidth of 30 nm, a repetition rate of 82 MHz, and an average
power of 300 mW. These pulses were sent to seed a regenerative
chirped pulse amplifier (Coherent, RegA 9000) that produced
approximately 70 fs, 5 $\mu$J pulses with a repetition rate of 250
kHz and an average power of 1.25 W. Around ten percent of this
average power was used as the pump beam; the remaining power was
directed to an optical parametric amplifier (Coherent, OPA 9400)
which generated a 4 mW average power Stokes beam, tunable from
400-1200 nm \cite{bredfeldt2003}.

\begin{figure}[htbp]
\centering\includegraphics[width=9cm]{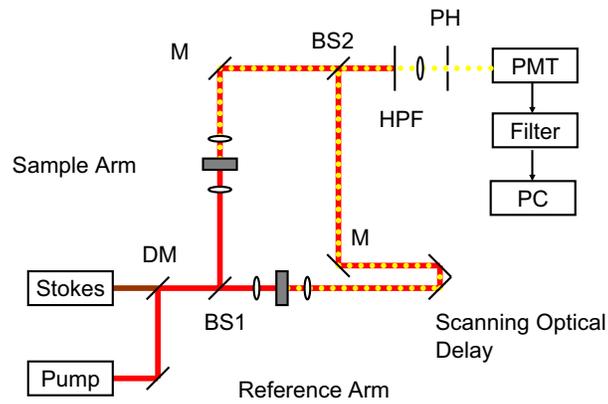} \caption{Setup of
the interferometric CARS measurement system
\cite{marks2003bis,bredfeldt2003}.  DM, dichroic mirror; BS,
beamsplitter; M, mirror; HPF, high-pass-filter; PH, pin-hole; PMT
photomultiplier tube; PC, personal computer. } \label{setupcars}
\end{figure}

Once the pump and the Stokes fields were generated, the two fields
entered the interferometer shown in detail in Fig.\ref{setupcars}.
%
%
 An excitation field consisting of two
overlapped pulses centered at the pump and Stokes wavelengths was
divided by a beamsplitter into two separate interferometer paths,
which are referred to in Fig.\ref{setupcars} as  ``sample arm''
and ``reference arm''.

A sample of a molecule was placed into each arm into which the
split excitation fields were focused. When the Stokes pulses were
tuned such that the difference in  frequency between the pump and
Stokes pulses was equal to a Raman active vibrational mode present
in the molecule in both the sample arm and the reference arm, an
anti-Stokes signal was generated in each arm of the
interferometer. It will follow that the anti-Stokes fields, being
coherent with the incident fields, will interfere at the
beamsplitter BS2 when temporally and spatially overlapped.

The pump and Stokes pulses, at 807 and 1072 nm respectively, were
used to excite the Raman-active vibrational mode of benzene at
3063 cm$^{-1}$. The pulses were collinearly overlapped using a
dichroic mirror and split with a 50:50 ultrafast beamsplitter
(BS1) into the sample arm  and the reference arm. The positions of
the samples in the sample and reference arms were chosen such that
the distances of the samples from the beamsplitter BS1 were equal.
The two generated anti-Stokes pulses were then overlapped in time
by adjusting the relative delay (delay line) and in space by
adjusting the position on a second beamsplitter (BS2). The
anti-Stokes signal was spectrally and spatially filtered. The
delay line in the reference arm was scanned by a
computer-controlled single axis translation stage at a constant
rate. The CARS signal intensity was measured with a
photomultiplier tube (PMT). The signal was filtered with a
low-pass anti-aliasing filter and sampled with a PC based data
acquisition system (NI DAQ, National Instruments).

\subsection{Results and discussion}
\begin{figure}[htbp]
\centering\includegraphics[width=9cm,angle=270]{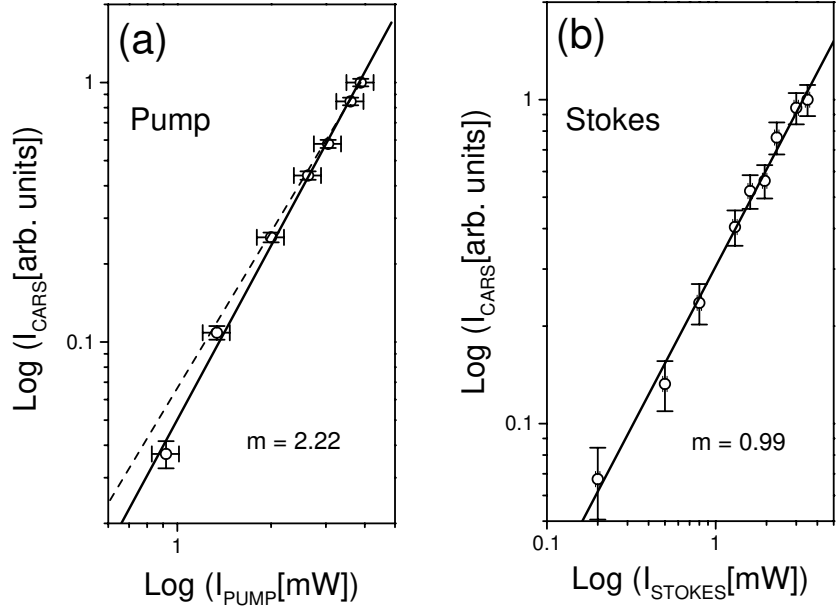}
\caption{ Log-log plots of the intensity of the CARS signal as a
function of (a) the intensity of the Pump field and (b) the
intensity of the Stokes field (solid lines, curve fitting). The
dotted line in (a) has a slope of 2. ``m'' is the angular
coefficient of the solid lines \cite{bredfeldt2003}.}
\label{squarelaw}
\end{figure}

Initially the interferometer was calibrated with the pump signal
only ($\lambda$ = 807 nm). The acquired interferogram
(Fig.\ref{interfpump}) is shown for reference as a comparison with
the following interferograms shown below. The interferogram was
detected at the beamsplitter BS2 (Fig.\ref{setupcars}) and was
recorded as the pathlength of the reference arm was scanned.
%
%
\begin{figure}[htbp]
\centering\includegraphics[width=9cm]{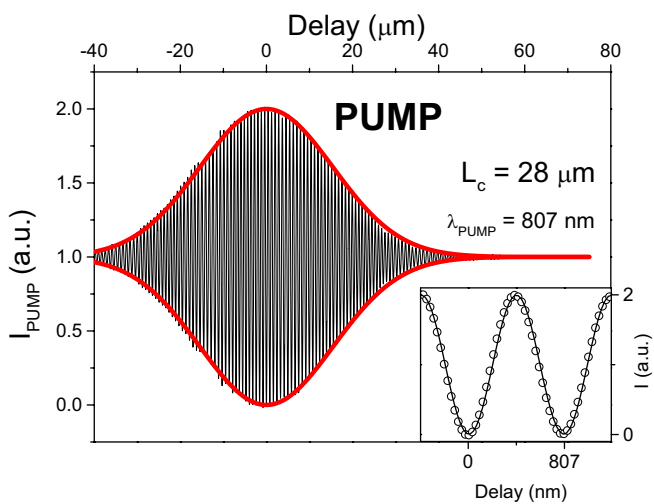}
\caption{Interferogram of the pump beam detected at the
beamsplitter BS2 of the setup shown in Fig.\ref{setupcars}. The
envelope of the interferogram was fitted using the modulus of the
degree of the coherence function. In the inset is shown a detail
of the interference pattern and its fit by the real part of the
degree of coherence function (open circles: experimental data;
solid line: fit). L$_c$ is the coherence length of the pulse.
$\lambda_{PUMP}$ is the wavelength of the Pump signal.}
\label{interfpump}
\end{figure}
Having determined the interferogram for the pump field, we
inserted in the two interferometer arms two quartz cuvettes filled
with benzene.  We first demonstrate that the signal collected from
the cuvette was a CARS signal. Figs. \ref{squarelaw}(a) and
\ref{squarelaw}(b) show the observed relationship between the CARS
and the pump intensity (Stokes intensity fixed) and the CARS and
the Stokes intensity (pump intensity fixed), respectively.
%
%
In agreement with the theory, the slope of the fitted lines verify
the linear relationship between the anti-Stokes and the Stokes
intensities, and the quadratic relationship between the
anti-Stokes and the pump intensities. This implies that our signal
is a result of a four-wave mixing process. In addition, we
observed that this process is CARS resonance because the
anti-Stokes power is maximized when the Stokes wavelength is tuned
to resonance with the Raman-active benzene vibrational mode.

Established that the filtered signals from both the cuvettes were
CARS signals, we detected the resulting signal at the beamsplitter
BS2 of the setup shown in Fig.\ref{setupcars}. The measured
interferogram results from the interference between the two
anti-Stokes signals and was recorded as the pathlength of the
reference arm was scanned.
%
%
\begin{figure}[htbp]
\centering\includegraphics[width=9cm]{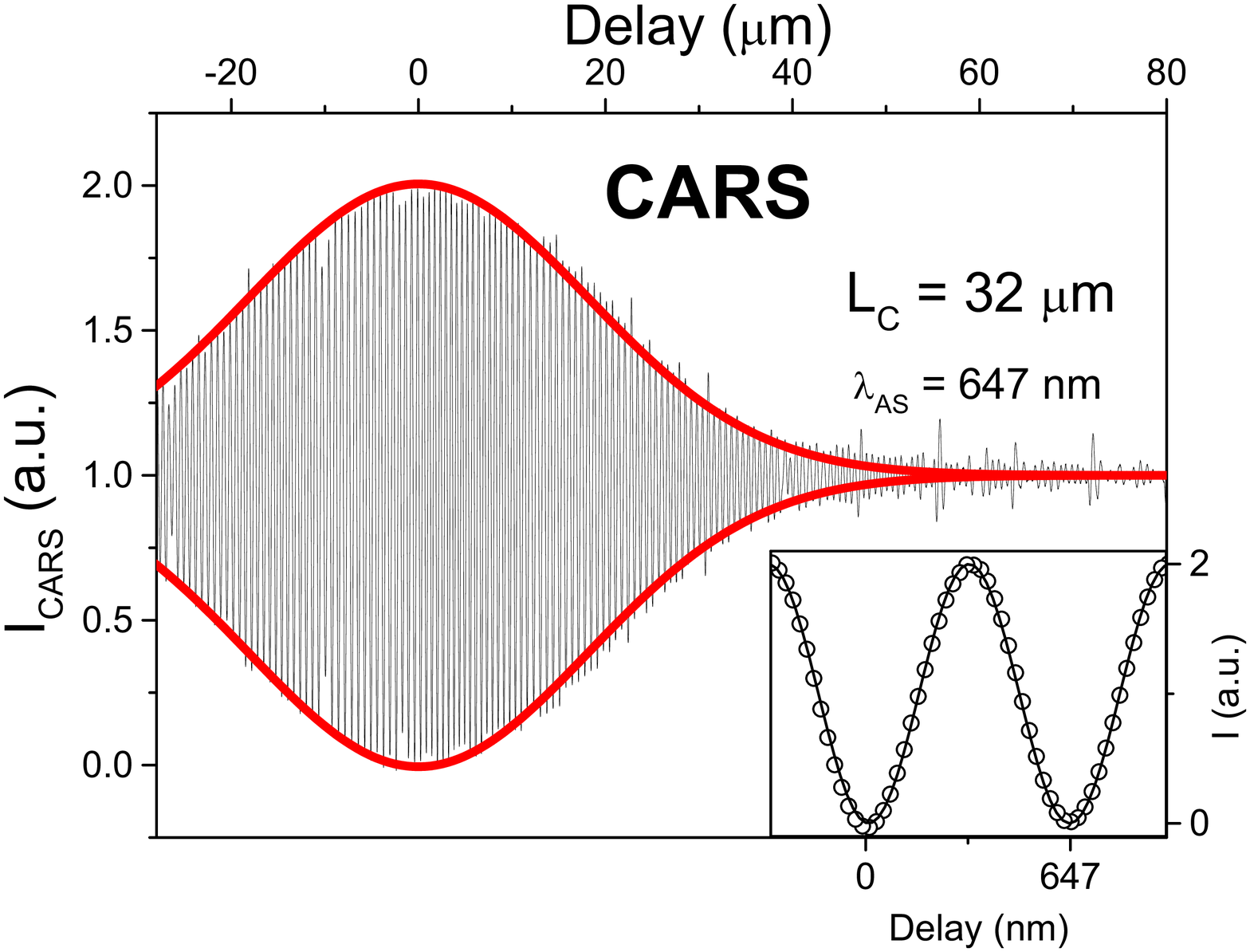} \caption{CARS
interferogram of the pump beam detected at the beamsplitter BS2 of
the setup shown in Fig.\ref{setupcars}. The envelope of the
interferogram was fitted using the modulus of the degree of the
coherence function. In the inset is shown a detail of the
interference pattern and its fit by the real part of the degree of
coherence function (open circles: experimental data; solid line:
fit). L$_c$ is the coherence length of the pulse. $\lambda_{AS}$
is the wavelength of the CARS signal.} \label{interfcars}
\end{figure}
The function used to fit the experimental data is the degree of
coherence function, that under the assumption of a Gaussian
spectral distribution, is given by
\begin{equation}
\gamma(\tau) = exp\left( -i\tau\omega_0 - i\frac{\delta \omega^2
\tau^2}{4}\right)
\end{equation}
where  $\tau$ is the time, $\omega_0$ is the center frequency and
$\delta \omega$ is the bandwidth of the CARS pulse. This function
is used to fit the interferogram in Fig.\ref{interfcars}. The real
part and the modulus of the coherence function, under the
assumption of a Gaussian spectral distribution, are used to fit
the experimental data (interferogram and envelope respectively)
and are plotted in Fig. \ref{interfcars}.

The resulting coherence length L$_C$,  defined as  the axial
resolution of the interferometric CARS measurement technique, is
equal to
\begin{equation}
\tau_c = \int^{+\infty}_{-\infty} |\gamma(\tau)|^2 d\tau
\end{equation}
and was found to be equal to 32 $\mu$m.

The possibility to demodulate interferometrically the two
anti-Stokes signals generated in separate samples demonstrates the
potential of CARS as a promising technique for providing molecular
contrast for OCT-like interferometric imaging systems.
Interference indicates  that similar Raman-active vibrational
frequencies were present in both the reference and the sample arm.
The ``fingerprint'' nature of Raman spectroscopy and the
possibility to switch between different molecular species in the
reference arm, could permit selective detection and imaging of
different molecules in the sample.

Moreover, the possibility to interfere a weak CARS signal with
another strong CARS signal (produced in the reference arm),
provides heterodyne sensitivity and an improved S/N ratio.

\section{SHG contrast enhancement}
\label{sectionshg}
\subsection{Theory}
Second Harmonic Generation (SHG), also known as frequency
doubling, has recently emerged as a valid imaging contrast
mechanism for microscopic imaging of cell and tissue structures
and functions \cite{campagnola2003}. As mentioned in Sec.
\ref{carstheory} SHG, in contrast to CARS, is a $\chi^{(2)}$
process in which two photons at the fundamental frequency are
converted into a single photon at exactly twice the frequency
without having any absorption and/or re-emission from the sample
\cite{stoller2003}. Even in this case the intensity of the
incident light is responsible for the induced nonlinear
polarization, with the result that the amplitude of the SHG signal
is proportional to the square of the incident light intensity.

The first biological SHG imaging experiment \cite{campagnola2003}
dates back to 1986 and involved the study of orientation of
collagen fibers in rat tail tendon \cite{freund1986}. Since then,
SHG microscopy has been successfully applied in many different
fields. In particular, SHG has proved to be highly effective in
selectively probing interfaces, without being overwhelmed by the
signal coming from the  bulk media \cite{eisenthal1996}. The
reason for this
 is that second-order processes are electric dipole
forbidden in centrosymmetric media. This implies bulk liquids and
centrosymmetric crystalline solids do not generate second-harmonic
signals. Instead, at the interfaces, the molecular and atomic
species experience different interactions and the inversion
symmetry, which is present in the bulk, is broken
\cite{eisenthal1996}. Another advantage of SHG microscopy is the
high resolution typically achieved in nonlinear microscopy, and
its applications for imaging structures in live tissues consisting
of endogenous proteins such as collagen. Note that the contrast
mechanism is obtained without requiring the presence of any
exogenous labels \cite{brown2003}.

For all these reasons, SHG microscopy is a good candidate for
providing contrast enhancement for OCT. In the next section we
will demonstrate that the interferometer presented in Sec.
\ref{carstheory} can be analogously used for SHG heterodyne
detection due to the fact that SHG, like CARS, is a process
coherent with the excitation field.

\subsection{Experimental Setup}
\begin{figure}[htbp]
\centering\includegraphics[width=9cm]{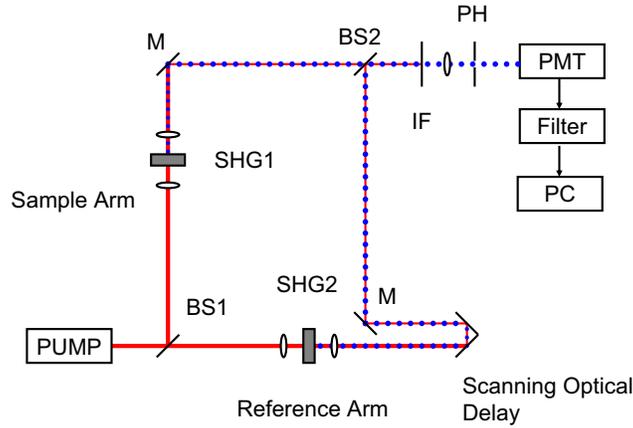} \caption{Setup of
the interferometric SHG measurement system. Two different SHG
crystals (Type I) were inserted in the two arms of the
interferometers. BS, beamsplitter; M, mirror; IF, interference
filter; PH, pin-hole; PMT, photomultiplier tube; PC, personal
computer. } \label{setupshg}
\end{figure}

The SHG interferometer is similar to the CARS interferometer and
is shown in Fig.\ref{setupshg}.
%
%
In this configuration, instead of having a Stokes and a Pump
laser, only the Pump laser at 807 nm was present. In the reference
arm, a reference nonlinear crystal (BBO, Type I) with a thickness
of 100 $\mu$m was present in which SHG signal was created. In the
sample arm, a different  nonlinear crystal (BBO, Type I) with a
thickness of 1 mm, was present. The signals generated in both the
crystals were overlapped as in the CARS configuration scheme.

Unique to the general methodology of our technique, when a SHG
crystal is placed in the reference arm, SHG signal created in the
sample under investigation and present in the sample arm can be
heterodyne-detected, allowing for high sensitivity detection and
OCT imaging.

\subsection{Results and discussion}
\begin{figure}[htbp]
\centering\includegraphics[width=9cm]{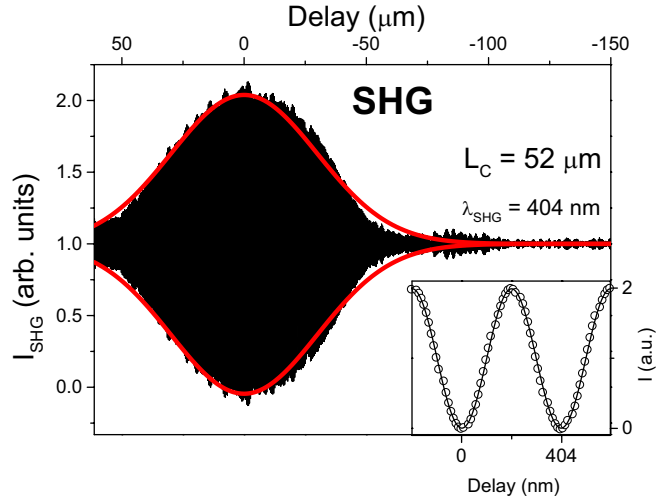} \caption{SHG
interferogram detected at the beamsplitter BS2 of the setup shown
in Fig.\ref{setupshg}. The interferogram was recorded as the
pathlength of the reference arm was scanned. The modulus of the
degree of the coherence function was used to fit the envelope of
the interferogram. The inset shows a detail of the interference
pattern and its fit by the real part of the degree of coherence
function (open circles: experimental data; solid line: fit). L$_c$
is the coherence length of the pulse.  SHG is the wavelength of
the SHG signal.} \label{interfshg}
\end{figure}

Fig.\ref{interfshg} shows the measured interferogram resulting
from the interference between the two SHG signals.
%
%
This result indicates that two SHG signals generated in separate
samples using the same pump laser can be demodulated
interferometrically. In this case as well, the presence of
interference clearly demonstrates the potential of SHG and
resonance-enhanced SHG as a promising technique for providing
molecular contrast for OCT-like interferometric imaging systems.
The presence of a nonlinear crystal in the reference arm will
allow one to interferometrically demodulate the SHG signal created
in the sample under investigation.

\section{Simulations}

Simulations of the coherent nonlinear processes were performed to
determine the properties of the signals that could be expected
from both the nonlinear interferometric setups
(Fig.\ref{setupcars}, and Fig.\ref{setupshg}) utilizing SHG and
CARS. Because of the wide bandwidths typically used in OCT
signals, the slowly varying envelope approximation did not suffice
for these simulations. Instead, we utilized a method that
approximated a continuous signal by sampling it spatially at
regular intervals. The evolution of the nonlinear signal was
achieved by applying the nonlinearity at each point in space and
time, and propagating the signal forward in time in uniform steps.
Because of the low conversion efficiency, we assumed that the
incident wave would not change in time except for linear
propagation effects, and so would experience no depletion in
energy. The model assumed that the incident and excited waves were
one-dimensional plane waves.  As a result, walk-off effects
resulting from the divergence of the wave and ray (Poynting)
vectors that occur with a finite incident beam size in a
birefringent medium were not accounted for.
%
%
\begin{figure}[htbp]
\centering\includegraphics[width=11cm]{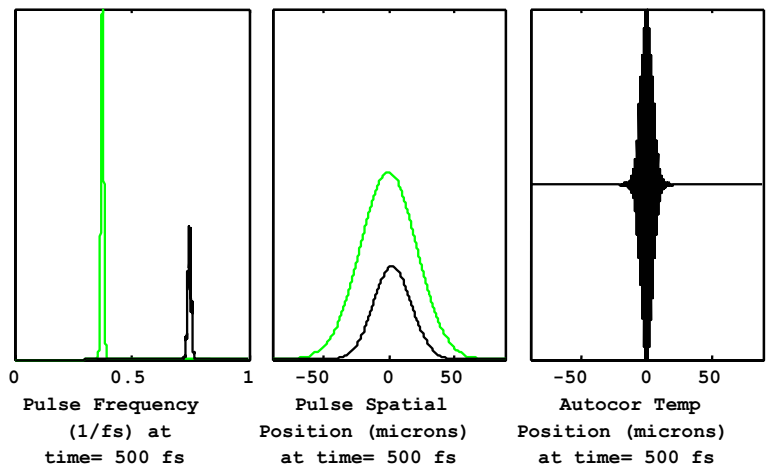}
\caption{Simulation of SHG. Click on figure to view an mpeg movie
(817 kB) of nonlinear interferometry of the coherent SHG process.
} \label{simshg}
\end{figure}
For the simulation of SHG, the nonlinearity was assumed to be
instantaneous, but the dispersion of the BBO crystal was accounted
for, so that phase matching effects (at 28 degrees inclination of
the extraordinary axis) were properly simulated. The incoming wave
was negatively chirped, with an 807 nm center frequency and 30 nm
bandwidth.  Because of the rather thick crystal (1 mm) used in the
experiment and present in the sample arm, 5 ps of simulation time
was used to interact the waves. Due to the thickness, phase
matching can not be achieved over the entire incident bandwidth.
The results of the simulation are shown in  the first movie, a
clip of which is shown in Fig.\ref{simshg}. The left graph shows
the  temporal spectrum of the incident beam in green, and the
second harmonic spectrum in black as the pulse evolves in time.
The middle graph shows the spatial envelope of the incident pulse
in green and the created second harmonic pulse in black. Finally,
the right graph shows the autocorrelation of the second harmonic
pulse at that time, which corresponds to the interferogram that
would be measured from two identical BBO crystals of a thickness
that corresponds to the propagation time of the simulation.  The
spatial origin of the simulation moves with the reference frame of
the incident pulse.

For the first 500 fs, the bandwidth is approximately $\sqrt{2}$
times that of the incident pulse.  As phase mismatch and group
velocity mismatch start to occur, the nonlinear polarization adds
destructively with the second harmonic pulse.  The resulting
spectrum of the pulse becomes quite nonuniform.  Because the
excited signal is simulated as a plane wave, all of the generated
frequency components overlap spatially and contribute to the
autocorrelation of the measured interference signal.  Therefore
the simulated autocorrelation has many sidelobes corresponding to
the nonuniform spectrum.  In the experimental setup, a focused
beam was used, and so the walk-off separated the spectral
components of the second harmonic pulse spatially.  At the pinhole
PH present at the exit of the interferometer (Fig.\ref{setupshg}),
only a limited bandwidth could be focused through, and so the
autocorrelation length was much longer than would have been
expected if phase matching occurred over the entire incident pulse
bandwidth. We expect that with a shorter crystal, phase-matching
could be better achieved and the autocorrelation could achieve its
theoretical minimum for the bandwidth available from the incident
pulse.

Simulating the CARS process require a different approach  because
unlike a nonresonant nonlinear second-harmonic generation process,
a resonant process has a ``memory'' associated with the nonlinear
polarization of the Raman process at each point in the medium.
Walk-off did not occur because the simulated medium, liquid
benzene, is isotropic.  The incoming radiation was assumed to be
two overlapped pulses of 30 nm bandwidth at 807 and 1072 nm, which
created a beat frequency at a vibrational frequency of the
benzene. The vibration was assumed to be a single vibration of
Lorentzian profile with a much longer lifetime than the duration
of the incident pulse.  At each time step the nonlinear Raman
polarization at each point in the medium was updated by driving it
with the instantaneous intensity of the incident pulse at that
point.  Fluctuations in the intensity of the incident field at the
Raman frequency will excite the nonlinear polarization at that
point in the medium.  This nonlinear polarization was mixed with
the incident beam to create the anti-Stokes signal.
%
%
\begin{figure}[htbp]
\centering\includegraphics[width=11cm]{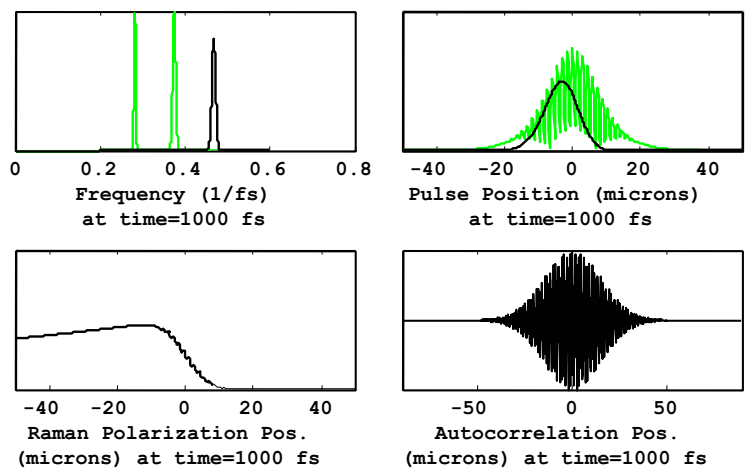}
\caption{Simulation of CARS. Click on figure to view an mpeg movie
(1.713 Mb) of nonlinear interferometry of the coherent CARS
process. } \label{simcars}
\end{figure}
The second movie, shown in Fig.\ref{simcars}, details the
evolution of the simulation of the anti-Stokes pulse in liquid
benzene over 1 ps. The upper left graph shows the temporal
spectrum of the incident pulse in green, and the created
anti-Stokes spectrum in black. The upper right graph shows the
spatial profile of the magnitude of the incident pulse in green,
and the magnitude of the created anti-Stokes pulse in black. The
lower left shows the spatial distribution of the magnitude of the
Raman polarization of the benzene medium. Finally, the lower right
shows the autocorrelation of the anti-Stokes pulse. Initially, as
the pulse propagates the medium is not excited, and therefore no
anti-Stokes light can be created. As the pulse moves through a
given point in the medium, the medium is resonantly excited.  The
medium then mixes with the incoming pulse to generate an
anti-Stokes wave on the trailing edge of the incident pulse, where
the maximum Raman polarization coincides with the incident pulse.
Because the anti-Stokes and incident pulses are close in
frequency, and dispersion is low in the near infrared, the
anti-Stokes pulse can remain in phase with the incident pulse for
a long distance in the medium and phase matching is easier to
achieve.  Because of the lack of walk-off, the simulation predicts
well the autocorrelation observed by experiment.

\section{Conclusion}
In conclusion, we have described a novel technique for contrast
enhancement in OCT based on optical nonlinearities.  The contrast
mechanisms are based on resonant enhancement of the third order
and second order nonlinear susceptibility of the molecules under
investigation. The interference between two CARS signals generated
in separate samples (or alternatively two SHG signals), was
observed, allowing for heterodyne detection. Numerical simulations
for both coherent nonlinear processes were performed in order to
determine the properties of the signal expected at the exit of the
described nonlinear interferometers, and the predicted results are
in agreement with the experimental data. The proposed
interferometric scheme is very promising for the development of a
new molecular imaging technique (NIVI) based on nonlinear,
low-coherence interferometry \cite{marks2003,bredfeldt2003} and
for SHG-OCT.

In this work, we focused on forward CARS and SHG, but epi-detected
CARS and SHG are coherent as well and are compatible with OCT
coherence-ranging systems. CARS and SHG interferometry provide the
advantages of interferometric detection and at the same time
provide OCT with molecular-specific contrast. These advantages
could make CARS and SHG interferometry a powerful tool for
biological imaging with OCT. Moreover, the same configuration
scheme could be exploited for Third Harmonic Generation (THG)
microscopy \cite{squier1998}, Coherent Stokes Raman Scattering
(CSRS) microscopy, and other coherent scattering processes.

\section{Acknowledgements}
This research was supported in part by a research grant entitled
"A Nonlinear OCT System for Biomolecular Detection and
Intervention" from NASA and the National Cancer Institute
(NAS2-02057, SAB). S.A. Boppart's email address is
boppart@uiuc.edu.

%
%

\end{document}